\documentclass[prl,showpacs]{revtex4}
\usepackage{graphics}
\usepackage{epsfig}
\usepackage{graphicx}

\usepackage{amsfonts} % packet for \mathbb{R}

\newcommand{\ba}{\begin{eqnarray}}
\newcommand{\ea}{\end{eqnarray}}
\newcommand{\ua}{\uparrow}
\newcommand{\da}{\downarrow}

\begin{document}

\title{Fidelity Between Partial States as Signature of Quantum Phase Transitions}

\author{N. Paunkovi\'c}
\affiliation{SQIG -- Instituto de Telecomunica\c{c}\~oes, IST,
 P-1049-001 Lisbon, Portugal}

\author{P. D. Sacramento, P. Nogueira and V. R. Vieira}
\affiliation{CFIF and Department of Physics, IST, Technical
University of Lisbon (TULisbon), P-1049-001 Lisbon, Portugal}

%\author{Paulo Nogueira}
%\affiliation{CFIF and Department of Physics, IST, Technical
%University of Lisbon, P-1049-001 Lisbon, Portugal}

%\author{Vitor Rocha Vieira}
%\affiliation{CFIF and Department of Physics, IST, Technical
%University of Lisbon, P-1049-001 Lisbon, Portugal}

\author{V. K. Dugaev}
\affiliation{Department of Mathematics and Applied Physics,
Rsesz\'ow University of Technology, Al. Powsta\'nc\'ow Warszawy 6,
35-959 Rsesz\'ow, Poland}

\date{\today}

\begin{abstract}
We introduce a partial state fidelity approach to quantum phase
transitions.
We consider a superconducting lattice with a magnetic impurity
inserted at its centre, and look at the fidelity between partial
(either one-site or two-site) quantum states.
In the vicinity of the point of the
quantum phase transition, we observe a sudden drop of the fidelity
between two one-site partial states corresponding to the
impurity location and its close vicinity. In the case of
two-site states, the fidelity reveals the transition point as long as
one of the two electron sites is located at the impurity, while the other
lies elsewhere in the lattice. We also determine the Uhlmann
mixed state geometric phase, recently introduced in the study of the
structural change of the system state eigenvectors in the vicinity
of the lines of thermal phase transitions, and find it to be trivial,
both for one- and two-site partial states, except when an electron
site is at the impurity.
This means that the system partial state eigenvectors do not contribute
significantly to the enhanced state distinguishability around the
point of this quantum phase transition. Finally, we use the fidelity
to analyze the total amount of correlations contained within a
composite system, showing that, even for the smallest two-site
states, it features an abrupt quantitative change in the vicinity of
the point of the quantum phase transition.
\end{abstract}

\maketitle

Recently, considerable developments have been made within the
research born out of the collaboration between physicists from two
rather distinct fields, viz those of quantum information and
computation and those of condensed matter and many-body physics.
Most of this interdisciplinary work has focused on the realization
of quantum information protocols with real many-body systems (for an
overview, see \cite{nielsen}). Yet, a considerable development has
been accomplished by moving in the ``opposite'' direction, ie
applying concepts and techniques widely used within the field of
quantum information and computation to the study of problems
relevant within the field of condensed matter physics. One of such
topics, recently researched to a great extent, is that of the use of
entanglement in the study of zero-temperature quantum phase
transitions \cite{sachdev}, as well as general thermal phase
transitions (for an overview, see \cite{us-book}, chapters
$19$--$21$, and references therein).

One application of the quantum information techniques to
problems dealing with macroscopic many-body systems is the use of
the (ground) state fidelity \cite{wootters} in studying (quantum)
phase transitions. It was first reported in \cite{zanardi-first},
where the sudden drop of the fidelity along the regions of quantum
phase transitions was observed in the Dicke and the
$XY$ models. Consequently, this method was applied to the cases of
free Fermi systems \cite{zanardi-free_fermion}, Bose-Hubbard
models \cite{buonsante-prl,oelkers}, spin chains
\cite{chen-excited,min} and in connecting the fidelity with the
renormalization group flows \cite{zhou}. Also, the cases of
non-symmetry breaking quantum phase transitions that are not of the
Landau-Ginzburg-Wilson type were studied
\cite{zanardi-matrix_prod,hamma-topological_order,gu-kostrelitz}.
A more formal and general study based on differential
geometric insights, establishing a connection between quantum phase
transitions and the relevant thermodynamic response functions, was
developed in \cite{zanardi-differential,zanardi-scaling} (see also
\cite{wen-long-thermal}). Finally, using the mixed state fidelity,
and extending the parameter space to include the temperature,
quantum phase transitions were studied in \cite{zanardi-thermal},
while the thermal phase transitions were explored in
\cite{zanardi-differential,wen-long-thermal,nikola-vitor}. It is
also worth noting here that the recently studied connection between
the quantum phase transitions and Berry geometric phases
\cite{qpt-berry} was established in
\cite{zanardi-differential,zanardi-scaling} on a more formal level,
unifying the fidelity and the Berry phase approaches to quantum
phase transitions. Subsequently, the Uhlmann mixed state geometric
phase \cite{uhlmann} (the generalization of the pure state Berry
phase) was introduced in \cite{nikola-vitor} as a tool in exploring
the change of the system thermal equilibrium state eigenvectors in
the vicinity of the line of a phase transition.

Given two physical systems $a$ and $b$, their quantum states, being
either pure or mixed, are given by density operators $\hat{\rho}_a$
and $\hat{\rho}_b$. Since according to quantum mechanics it is in
general not possible to unambiguously distinguish with probability
one between two given states \cite{nielsen, fuchs}, various measures
of quantum state distinguishability have been introduced (for an
overview, see \cite{fuchs}). One of the most widely used measures is
(quantum) fidelity \cite{wootters}, given by the expression
\begin{equation}
\label{fidelity-def} F(\hat{\rho}_a, \hat{\rho}_b) = \mbox{Tr}
\sqrt{\sqrt{\hat{\rho}_a}\hat{\rho}_b\sqrt{\hat{\rho}_a}}.
\end{equation}
The numerical value of the fidelity ranges from $F=0$, in the case
of fully distinguishable states \footnote{Note that in the case of
pure states, this corresponds to the case of mutually orthogonal
state vectors.}, to $F=1$, when the two quantum states are
completely indistinguishable from each other (ie they are
identical). As it happened with most other quantum state
distinguishability measures, the introduction of the
quantum fidelity $F$ was also motivated by the analogous classical
quantity. Classical fidelity $F_c$, a measure of distinguishability
between two probability distributions $\{p_a(i)\}$ and $\{p_b(i)\}$,
is defined as $F_c(p_a,p_b)=\sum_i\sqrt{p_a(i)p_b(i)}$. Measuring an
observable $\hat{A}$ in states $\hat{\rho}_a$ and $\hat{\rho}_b$,
one obtains probability distributions $\{p^{\hat{A}}_a(i)\}$ and
$\{p^{\hat{A}}_b(i)\})$ respectively, whose mutual
distinguishability is given by the classical fidelity
$F_c(p^{\hat{A}}_a,p^{\hat{A}}_b)$. It can be proven (see
\cite{fuchs}) that for any observable $\hat{A}$, the
inequality $F(\hat{\rho}_a, \hat{\rho}_b)\leq
F_c(p^{\hat{A}}_a,p^{\hat{A}}_b)$ holds, and there always exists an
optimal observable $\hat{A}_{op}$ for which the equality is
achieved. In other words, quantum fidelity $F(\hat{\rho}_a,
\hat{\rho}_b)$ gives the optimal value for the distinguishability
between two quantum states, when compared through probability
distributions $\{p^{\hat{A}}_a(i)\})$ and $\{p^{\hat{A}}_b(i)\}$
(note though that it does not give an optimal observable
$\hat{A}_{op}$). Using the modulus of an operator $\hat{R}$,
$|\hat{R}|=(\hat{R}\hat{R}^\dag)^{1/2}$, one obtains an alternative
expression for the fidelity \cite{nielsen}:
\begin{equation}
\label{fidelity-alternative}
 F(\hat{\rho}_a, \hat{\rho}_b)= \mbox{Tr}
 \left|\sqrt{\hat{\rho}_a}\sqrt{\hat{\rho}_b}\,\right| =
 \mbox{Tr}\sqrt{\sqrt{\hat{\rho}_a}\sqrt{\hat{\rho}_b}(\sqrt{\hat{\rho}_a}\sqrt{\hat{\rho}_b})^\dag}
% = \mbox{Tr}\sqrt{\sqrt{\hat{\rho}_a}\sqrt{\hat{\rho}_b}\sqrt{\hat{\rho}_b}\sqrt{\hat{\rho}_a}}
 = \mbox{Tr}\sqrt{\sqrt{\hat{\rho}_a}\hat{\rho}_b\sqrt{\hat{\rho}_a}}.
\end{equation}

The fidelity approach to quantum phase transitions is based on the
use of the above quantity (\ref{fidelity-def}) in distinguishing
macroscopic quantum states of many-body quantum systems. Consider a system
given by the set of Hamiltonians $\hat{H}(q)$, where $q$ is a
parameter representing a set of interaction coupling constants, which at
$T=0$ is in a pure ground state $|\Phi\rangle=|\Phi(q)\rangle$.
Then, the sudden drop of the fidelity (\ref{fidelity-def})
$F(|\Phi(q)\rangle\langle\Phi(q)|,|\Phi(\tilde{q})\rangle\langle\Phi(\tilde{q})|)=|\langle\Phi(q)|\Phi(\tilde{q})\rangle|$
between two ground states $|\Phi(q)\rangle$ and
$|\Phi(\tilde{q})\rangle$ obtained for two close parameter points
$q$ and $\tilde{q}=q+\delta q$, with $\delta q$ being small, can
indicate the points of quantum phase transitions. The basic logic
behind this idea is strikingly simple --- two quantum states defining
different macroscopic phases are expected to have enhanced
distinguishability that would quantitatively exceed that taken
between the states from the same phase. In our study, instead of the
global ground states that describe the overall system, we will study
 partial quantum states of certain subsystems of the
system. If the overall system $\mathcal{S}$ is divided into two
parts $A$ and $B$, then we will study the mixed state fidelity
$F(\hat{\rho}_{A}(q),\hat{\rho}_{A}(\tilde{q}))$, where
$\hat{\rho}_{A}(q)=\mbox{Tr}_{B}|\Phi(q)\rangle\langle\Phi(q)|$, and
analogously for $\hat{\rho}_{A}(\tilde{q})$. We
will show that in the system considered, a conventional BCS
superconductor with an inserted magnetic impurity, a sudden drop of
the above mixed state fidelity can define the point at which the
(first order) quantum phase transition occurs, even when applied to
the smallest subsystems of the overall system.

\section{The Model}

In this work we consider the quantum phase transition induced by a
magnetic impurity inserted in a conventional superconductor
\cite{sakurai,satori,salkola,morr1,morr2,review,abrikosov60,ramazashvili97}.
Consider a classical spin immersed in a two-dimensional $s$-wave
conventional superconductor. We use a lattice description of the
system. In the centre of the system, $i=l_c=(x_c,y_c)$, we place a
classical spin parametrized by
\begin{equation}
\frac{\vec{S}}{S} = \cos\varphi\vec{e}_x+\sin\varphi\vec{e}_z.
\end{equation}

The Hamiltonian of the system is given by:
\begin{equation}
\label{Hamiltonian-def} \hat{H} = -\sum_{\langle ij\rangle\sigma}
t_{ij}\hat{c}_{i\sigma}^\dag\hat{c}_{j\sigma}-\varepsilon_F\sum_{i\sigma}\hat{c}_{i\sigma}^\dag\hat{c}_{i\sigma}
+\sum_i\left(\Delta_i\hat{c}_{i\ua}^\dag\hat{c}_{i\da}^\dag+\Delta_i^\ast\hat{c}_{i\da}\hat{c}_{i\ua}\right)
-\sum_{\sigma\sigma^\prime}
J\left[\cos\varphi\hat{c}_{l_c\sigma}^\dag\sigma^x_{\sigma\sigma^\prime}\hat{c}_{l_c\sigma^\prime}
+\sin\varphi\hat{c}_{l_c\sigma}^\dag\sigma^z_{\sigma\sigma^\prime}\hat{c}_{l_c\sigma^\prime}\right].
\end{equation}
where the first term describes the hopping of electrons between
different sites on the lattice, $\varepsilon_F$ is  the chemical
potential (equal to the Fermi energy at $T=0$), the third term is
the superconducting $s$-pairing with the site-dependent order
parameter $\Delta_i$, and the last term with $J>0$ is the exchange interaction
of an electron at site $i=l_c$ with the magnetic impurity. The
hopping matrix is given by $t_{ij}=t\delta_{j,i+\delta}$ where
$\delta$ is a vector to a nearest-neighbor site. Note that both the
indices $l$ and $i,j\in\{1,2,\ldots N\}$ specify sites on a
two-dimensional system ($N$ is the number of sites). We will take
energy units in terms of $t$ ($t=1$), and $\varepsilon_F=-1$.

Since the impurity spin acts like a local magnetic field the
electronic spin density will align along the local spin. For small
values of the coupling there is a negative spin density around the
impurity site. At the impurity site it is positive, as expected. For
larger couplings the spin density in the vicinity of the impurity
site is positive. At small couplings the many-body system screens
the effect induced by the impurity inducing fluctuations that
compensate the effect of the local field in a way that the overall
magnetization vanishes. However, for a strong enough coupling the
many-body system becomes magnetized in a discontinuous fashion. One
interpretation is that if $J$ is strong enough the impurity breaks a
Cooper pair and captures one of the electrons, leaving the other
electron unpaired, and thus the overall electronic system becomes
polarized. The impurity induces a pair of bound states inside the
superconducting energy gap, one at positive energy (with respect to
the chemical potential), and another at a symmetric negative energy.
Even though the spectrum is symmetric the spectral weights of the
two energy levels are not the same and their spin content is also
distinct. The analysis of the local density of states (LDOS)
\cite{morr2,first} shows that for small coupling the lowest positive
energy level has only a contribution from spin $\uparrow$ and the
first level with negative energy (symmetric to the other level) has
only a contribution from spin $\downarrow$. The magnitude of the
spectral weight at the impurity site is different for the two
states. Considering a higher value for the coupling one finds that
the levels inside the gap approach the Fermi level. There is a
critical value of the coupling for which the two levels cross in a
discontinuous way such that it coincides with the emergence of a
finite overall magnetization. After the level crossing occurs,
the nature of the states changes. The positive energy bound state
has now only a contribution from the spin $\downarrow$ component and
vice-versa, the first negative energy state has only contribution
from the spin component $\uparrow$ --- hence as the level crossing
takes place the spin content changes.

The diagonalization of this Hamiltonian is performed using the
Bogoliubov-Valatin transformation in the form:
\begin{eqnarray}
\hat{c}_{i\ua} & = & \sum_n[u_n(i,\ua)\hat{\gamma}_n-v_n^\ast(i,\ua)\hat{\gamma}_n^\dag] \nonumber \\
\hat{c}_{i\da} & = &
\sum_n[u_n(i,\da)\hat{\gamma}_n+v_n^\ast(i,\da)\hat{\gamma}_n^\dag].
\end{eqnarray}
Here $n$ is a complete set of states, $u_n$ and $v_n$ are related to
the eigenfunctions of the Hamiltonian (\ref{Hamiltonian-def}), and
the new fermionic operators $\hat{\gamma}_n$ are the quasiparticle
operators. These are chosen such that in terms of the new operators:
\begin{equation}
\hat{H}=E_g+\sum_n\epsilon_n\hat{\gamma}_n^\dag\hat{\gamma}_n,
\end{equation}
where $E_g$ is the ground state energy and $\epsilon_n$ are the
excitation energies. As a consequence:
\begin{eqnarray}
\left[\hat{H},\hat{\gamma}_n\right] & = & -\epsilon_n\hat{\gamma}_n \nonumber \\
\left[\hat{H},\hat{\gamma}_n^\dag\right]  & = &
\epsilon_n\hat{\gamma}_n^\dag.
\end{eqnarray}
The coefficients $u_n(i,\sigma)$, $v_n(i,\sigma)$ can be obtained
solving the Bogoliubov - de Gennes (BdG) equations \cite{degennes}.
Defining the vector:
\begin{equation}
\psi_n(i) = \left(\begin{array}{c} u_n(i,\ua) \\ v_n(i,\da) \\
u_n(i,\da) \\ v_n(i,\ua) \end{array}\right).
\end{equation}
the BdG equations can be written as:
\begin{equation}
\mathcal{H}\psi_n(i)=\epsilon_n\psi_n(i),
\end{equation}
where the $\mathcal{H}$ matrix at site $i$ is given by:
\begin{equation}
\mathcal{H} = \left(\begin{array}{cccc}
-h-\varepsilon_F-J\sin\varphi & \Delta_i & -J\cos\varphi
& 0
\\ \Delta_i^\ast & h+\varepsilon_F-J\sin\varphi & 0 &
-J\cos\varphi \\ -J\cos\varphi & 0 &
-h-\varepsilon_F+J\sin\varphi & \Delta_i \\ 0 &
-J\cos\varphi & \Delta_i^\ast &
h+\varepsilon_F+J\sin\varphi
\end{array}\right),
\end{equation}
where $h=t\hat{s}_\delta$ with $\hat{s}_\delta f(i)=f(i+\delta)$.
The solution of these equations gives both energy eigenvalues and
eigenstates. The problem involves the diagonalization of a
$(4N)\times(4N)$ matrix. The solution of the BdG equations is
performed self-consistently imposing at each iteration that
\begin{equation}
\Delta_i = \frac{g}{2}[\langle
\hat{c}_{i\ua}\hat{c}_{i\da}\rangle-\langle
\hat{c}_{i\da}\hat{c}_{i\ua}\rangle],
\end{equation}
where $g$ is the effective attractive interaction between the
electrons. Using the canonical transformation this can be written as
\begin{equation}
\Delta_i = -g\sum_{n,0<\epsilon_n<\hbar\omega_D}\left\{ f_n(\epsilon_n)[
u_n(i,\ua)v_n^\ast(i,\da)+u_n(i,\da)v_n^\ast(i,\ua)]
-\frac{1}{2}[u_n(i,\ua)v_n^\ast(i,\da)+u_n(i,\da)v_n^\ast(i,\ua)]\right\}
\end{equation}
where $\omega_D$ is the Debye frequency, and $f_n(\epsilon_n)$ is
the Fermi function defined as usual like
\begin{equation}
f_n(\epsilon_n) = \frac{1}{e^{\epsilon_n/T}+1}
\end{equation}
where $T$ is the temperature. We note that the Bogoliubov - de
Gennes equations are invariant under the substitutions
$\epsilon_n\rightarrow -\epsilon_n$, $u(\ua)\rightarrow v(\ua)$,
$v(\ua)\rightarrow u(\ua)$, $v(\da)\rightarrow -u(\da)$,
$u(\da)\rightarrow - v(\da)$.

\section{Results}

As we have already noted, in this paper we focus on the fidelity
between the partial states rather than the overall ground states, as
it was the case in the previous studies
\cite{zanardi-first,zanardi-free_fermion,buonsante-prl,zhou,zanardi-matrix_prod,hamma-topological_order,gu-kostrelitz,zanardi-differential,zanardi-scaling,chen-excited,zanardi-thermal,wen-long-thermal,
nikola-vitor}. At $T=0$, the system is in the pure ground state
$|\Phi\rangle$ and has a density operator
$\hat{\rho}=|\Phi\rangle\langle\Phi |$. Thus, if we divide the whole
system in two subsystems, say $A$ and $B$, then the partial mixed
state, given by the reduced density operator $\hat{\rho}_A$ for the
subsystem $A$, is defined as
\begin{equation}
\hat{\rho}_A=\mbox{Tr}_B \hat{\rho},
\end{equation}
where $\mbox{Tr}_B[\cdot ]$ represents the partial trace evaluated
over the Hilbert space $\mathcal{H}_B$ of the subsystem $B$. In
particular, we will calculate the matrix elements of one-site and
two-site partial mixed states, given in general for different sites
$i$ and $(i,j)$ respectively, and different values of the coupling
constant $J$. In order to infer the point of the quantum phase
transition, we will calculate the fidelity between two one-site and
two two-site states, given for two close parameters $J$ and
$J+\delta J$. Further, to better understand the influence of the
quantum phase transition point on the structure of partial states
(and therefore, to a certain extent, the overall state as well), we
examine the fidelity between two different partial states, given for
the same values of the coupling constant $J$, but different sites,
for both cases of one- and two-site states.

In many-body systems a description in terms of quantum pure states
and wave functions as their representatives in a certain basis (e.g.
position, momentum, energy,...) is quite involved and the second
quantization is the natural way to perform any calculation. The
mixed states are however easily defined in terms of pure states and
their matrix elements in some basis, and this approach is indeed not
difficult to implement using Fock states. The matrix elements of the
density matrix are simply defined in terms of correlation functions
of the whole system. For instance, in the case of the single-site
partial mixed states it can be shown that, using local basis states
$\mathcal{B}=\{|0\rangle,|\!\ua\da\rangle\,|\!\ua\rangle,|\!\da\rangle\}$,
which denote the four possible states --- unoccupied, double
occupied, single occupied with an electron with spin up and single
occupied with an electron with spin down, respectively --- the
corresponding density matrix reads as
\begin{equation}
\label{one-site_state} \rho_i=\left(\begin{array}{cccc}
\langle(1-\hat{n}_\ua)(1-\hat{n}_\da)\rangle &
\langle\hat{c}_{\ua}^\dag\hat{c}_{\da}^\dag\rangle & 0 & 0 \\
\langle\hat{c}_{\da}\hat{c}_{\ua}\rangle &
\langle\hat{n}_\ua\hat{n}_\da\rangle & 0 & 0 \\ 0 & 0 &
\langle\hat{n}_\ua(1-\hat{n}_\da)\rangle &
\langle\hat{c}_{\da}^\dag\hat{c}_{\ua}\rangle \\ 0 & 0 &
\langle\hat{c}_{\ua}^\dag\hat{c}_{\da}\rangle &
\langle(1-\hat{n}_\ua)\hat{n}_\da\rangle
\end{array}\right)_i,
\end{equation}
where the index $i$ denotes the site.
The spin and the charge parts decouple. The spin part couples
the two spin orientations (single occupied states) and the charge
part couples the empty and doubly occupied states. The diagonal
terms of the matrix describe the number of empty sites, the number
of doubly occupied sites, the number of spin up sites and the number
of spin down sites, respectively. The sum of the diagonal terms is
equal to $1$ due to normalization. The matrix is easily diagonalized
and the fidelity (\ref{fidelity-def}) between two different one-site
states obtained straightforwardly.

The correlation functions are easily solved using the representation
of the electron operators in terms of the BdG quasiparticle
operators. Specifically, we can write the single-site density matrix
as
\begin{equation}
\hat{\rho} = \sum_{n,m} \rho_{nm} |n\rangle \langle m|
\end{equation}
where $|n\rangle,|m \rangle \in \mathcal{B}$ are the four basis
states described above. Consider now an operator $\hat{O}$ defined
in terms of creation and annihilation operators of the electrons,
and evaluate
\begin{equation}
\mbox{Tr} \{ \hat{O} \hat{\rho} \} = \sum_{n,m} \langle n| \hat{O}
|m \rangle \rho_{mn} .
\end{equation}
To determine the matrix element $\rho_{nm}$ of the density matrix it
is enough, by inspection, to find which operator $\hat{O}$ is such
that only the matrix element $\langle n| \hat{O} |m \rangle $ is
nonzero. This will tell us which correlation function gives which
matrix element of the density matrix. Alternatively, one can express
the operator $|m \rangle \langle n|$ in terms of the creation and
annihilation operators.
%This is indeed possible, and after a couple of cases we know easily how to
%choose the operator $O$ to give a specific matrix element.
The results for $\rho_A$ are given by equation (\ref{one-site_state}) above.

These matrix elements are defined in the subspace of the states of
one site. However, since we have integrated out all the other sites,
we may now replace the matrix elements by the matrix elements over
the entire system (direct product with the states of the remaining
$N-1$ sites).

Computing the fidelity between the two-site partial states is more
involved.
Given two sites, $i$ and $j$, the two-site basis states are given by
the direct product of the one-site basis sets that were defined earlier,
and the reduced density matrix is now a square matrix of order $16$.
However, exactly half of the $256$ matrix elements are null --- the
nonzero ones being those that correspond to either even-even or odd-odd
states, where the parity reflects the number of electrons in the sites
$i$ and $j$, respectively.
Hence those two sets of states decouple and the reduced density matrix
can be written as a block diagonal matrix where each submatrix is a
square matrix or order $8$.
Still, the calculation is rather complex and for that reason we used
the computer program FORM \cite{FORM} to obtain symbolic expressions
for the matrix elements involved.
Note that one of the correlation functions involves the double occupancies
at the two sites, and thus one should deal with products of up to eight
electron operators (where each electron operator is given by a sum of
two quasiparticle operators); also, the signs that follow from the
anti-commutation relations of fermionic operators must be taken into
account.

\begin{figure}
%\begin{picture}(300,225)
%\put(-5,0){\includegraphics[width=0.60\textwidth]{Fig1}}
%\end{picture}
\includegraphics[width=0.60\textwidth]{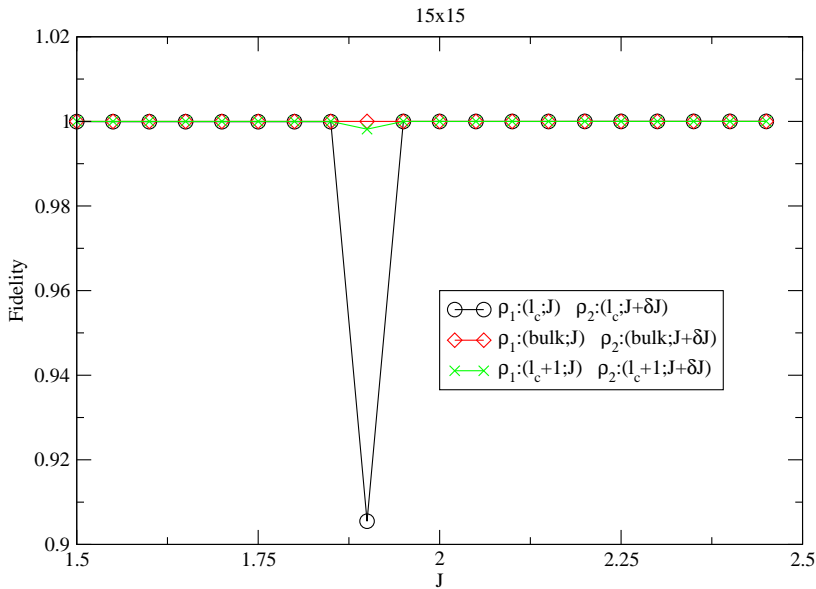}
\caption{\label{fig1} One-site fidelity between two states taken for
two close parameter points $J$ and $J+\delta J$ ($\delta J=0.05$),
for the site at the impurity (black line), a nearest neighbor (green
line) and a site located in the bulk (red line). Note the sudden
drop of the fidelity for the impurity around the point of the
quantum phase transition.}
\end{figure}

\subsection{Fidelity Between Single-site States}

In the following, we present the numerical results for the fidelity
between one-site partial states, for a two dimensional lattice of
size $15\times 15$ sites. For computational simplicity, we perform
our calculations for this system size. Since the effect of the
impurity is short-ranged, changing the size of the system only affects slightly the results.
First, we evaluate the fidelity (\ref{fidelity-def}) between two
one-site states given for two sites $i$ and $j$ and two close values
$J$ and $J+\delta J$ of the exchange parameter, with $\delta
J=0.05$, such that $\hat{\rho}_a=\hat{\rho}(J;i)$ and
$\hat{\rho}_b=\hat{\rho}(J+\delta J;j)$.

For $i=j=l_c$, when the electrons are located at the site of the
magnetic impurity, the results for the one-site fidelity
$F_1(\hat{\rho}_a,\hat{\rho}_b)$, as a function of $J$, are presented in Fig. \ref{fig1}.
We observe a clear drop, from the otherwise common value
$F_1(\hat{\rho}_a,\hat{\rho}_b)\simeq 1$, for the critical value
$J_0\simeq 1.9$ \cite{first} of the parameter $J$. As we move one of the electrons from the
impurity, the drop of the fidelity $F_1$ around the point of the
quantum phase transition becomes much smaller: for the first
neighbors of the impurity, it is still visible, while already for
the second-neighbors and the far away bulk sites, it becomes
negligible. We see that even for the smallest subsystems --- one-site
spins in the lattice --- the fidelity can be dramatically influenced
by the structure of their states, as long as they are close to the
impurity. This is intuitively easy to accept since it was shown in
the previous work \cite{first,impurity-entanglement}, through
investigation of various relevant physical quantities, such as the
gap, the electron density or the total and quantum correlations
(entanglement), that the impurity affects one-site properties of the system mainly in its
vicinity (although it was also shown that for some two-site entanglement
measures the effects persisted for larger distances).

\begin{figure}
%\begin{picture}(295,220)
%\put(-5,0){\includegraphics[width=0.60\textwidth]{Fig2}}
%\end{picture}
\includegraphics[width=0.60\textwidth]{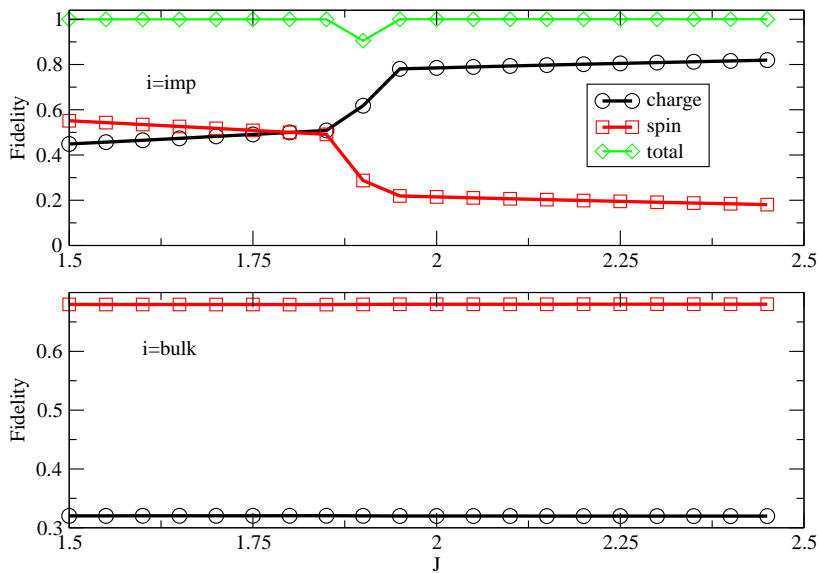}
\caption{\label{fig2} The charge and spin contributions of the
one-site fidelity as a function of the parameter $J$, with $\delta
J=0.05$, for the site at the  impurity (upper figure) and a site
located in the bulk (lower figure). Note the discontinuity for both
spin and charge contributions around the point of the quantum phase
transition, for the site located at the impurity.}
\end{figure}

A more detailed picture of the fidelity between the one-site states
is presented in Fig. \ref{fig2}, where the two distinct
contributions to the overall change of partial states is presented.
The charge and the spin parts of the one-particle state
(\ref{one-site_state}) lead to two different contributions to the
state distinguishability, and we see that they both exhibit a
discontinuity in the vicinity of the point of the quantum phase
transition.

Further, we have evaluated the Uhlmann mixed state geometric phase
\cite{uhlmann}, a generalization of the Berry geometric phase to
mixed quantum states. While the connection between Berry phases,
pure state fidelity and quantum phase transitions was studied
before~\cite{qpt-berry,zanardi-differential, zanardi-scaling}, the
Uhlmann mixed state geometric phase was recently introduced
\cite{nikola-vitor} to analyze the structural change of the system
eigenvectors, with respect to the change of the relevant parameter
driving the system Hamiltonian (ie $J$ in our case). The Uhlmann
geometric phase $\phi_g$ (and the corresponding holonomy) is
determined by the unitary operator $\hat{V}_{ab}$, given by the
polar decomposition (see for example \cite{nielsen})
$\sqrt{\hat{\rho}_a}\sqrt{\hat{\rho}_b}=|\sqrt{\hat{\rho}_a}\sqrt{\hat{\rho}_b}\,|
\hat{V}_{ab}$, where $|\hat{R}|=(\hat{R}\hat{R}^\dag)^{1/2}$
represents the modulus of the operator $\hat{R}$ (for details, see
\cite{sjoqvist}). Since the fidelity can be written as
$F(\hat{\rho}_a,\hat{\rho}_b)=
\mbox{Tr}|\sqrt{\hat{\rho}_a}\sqrt{\hat{\rho}_b}\,|$, see equation
(\ref{fidelity-alternative}), a nonzero difference
$H(\hat{\rho}_a,\hat{\rho}_b)-F(\hat{\rho}_a,\hat{\rho}_b)$, where
$H(\hat{\rho}_a,\hat{\rho}_b)=\mbox{Tr}[\sqrt{\hat{\rho}_a}\sqrt{\hat{\rho}_b}\,]$,
would imply $\hat{V}_{ab}\neq\hat{I}_{ab}$ and therefore the
emergence of a non-trivial Uhlmann geometric phase
$\phi_g\neq 0$. On the other hand, for the case of mutually
commuting Hamiltonians the system eigenvectors do not change with
the driving parameter and we have
$F(\hat{\rho}_a,\hat{\rho}_b)=\mbox{Tr}[\sqrt{\hat{\rho}_a}\sqrt{\hat{\rho}_b}\,]$.
In other words,
$F(\hat{\rho}_a,\hat{\rho}_b)=H(\hat{\rho}_a,\hat{\rho}_b)$ and the
Uhlmann geometric phase becomes trivial, ie $\phi_g=0$.

%$F_1(\hat{\rho}_a,\hat{\rho}_b)=\mbox{Tr}|\sqrt{\hat{\rho}_a}\sqrt{\hat{\rho}_b}| \simeq \mbox{Tr}\sqrt{\hat{\rho}_a}\sqrt{\hat{\rho}_b}$

For the one-site states, we find that the Uhlmann geometric phase is
always trivial for every site neighboring the impurity or in the
bulk: the one-site eigenvectors do not contribute significantly to
the change of the overall one-site partial states at the point of
the quantum phase transition. Still, for the impurity site itself,
around the point of the phase transition $J_0$ we observe a small
deviation from zero for the difference $H_1-F_1$, which indicates a
slight ``rotation'' of the electron's eigenbasis. This result can be
qualitatively explained noticing that in the model we are
considering the third term in the Hamiltonian
(\ref{Hamiltonian-def}), giving the superconducting features, and
the last term in the Hamiltonian, giving the coupling between the
impurity and the lattice electrons, commute with each other
\footnote{ These interaction terms are easily expressed in terms of
the electron spin operators $\hat{\vec{S}}$ and the Nambu operators
$\hat{\vec{T}}$, discussed in detail in \cite{nikola-vitor}. They
commute with each other, ie $[\hat{T}^\alpha,\hat{S}^\beta]=0$, for
every $\alpha,\beta\in\{0,+,-\}$, and they have a common eigenbasis.
In fact, these operators satisfy $\hat{T}^\alpha\hat{S}^\beta=0$,
and act non-trivially in mutually orthogonal subspaces, with even
and odd number of particle occupation numbers, respectively,
annihilating the states in the other subspace.}. Moreover, for a
{\em fixed} value of the gap $\Delta_i$, changing the
impurity-lattice coupling constant $J$, without changing the
impurity spin orientation angle $\varphi$, will result in no change
of the overall Hamiltonian's eigenbasis --- only the eigenvalues are
affected in this case and the system exhibits a first-order quantum
phase transition as a consequence of the explicit (not avoided)
level crossing. Thus, one might expect that also the partial
subsystem states would feature little change of local eigenbasis,
even in the vicinity of the point of the quantum phase transition.
This is precisely the case in the model considered, {\em apart} for
the gap $\Delta_{l_c}$ given for the electron located at the
impurity (see \cite{first}): for every site $i$, the gap $\Delta_i$
is a slowly changing function of $J$, while for the impurity
$\Delta_{l_c}$ exhibits a $\pi$ shift and becomes negative. Such a
behavior of $\Delta_{l_c}$ does affect the difference $H_1-F_1$, but
the effect is an order of magnitude smaller than for the fidelity.
Further, it affects the impurity site only and thus does not lead to
the superconductor-normal metal type of phase transition.

\begin{figure*}
\includegraphics[width=0.4\textwidth]{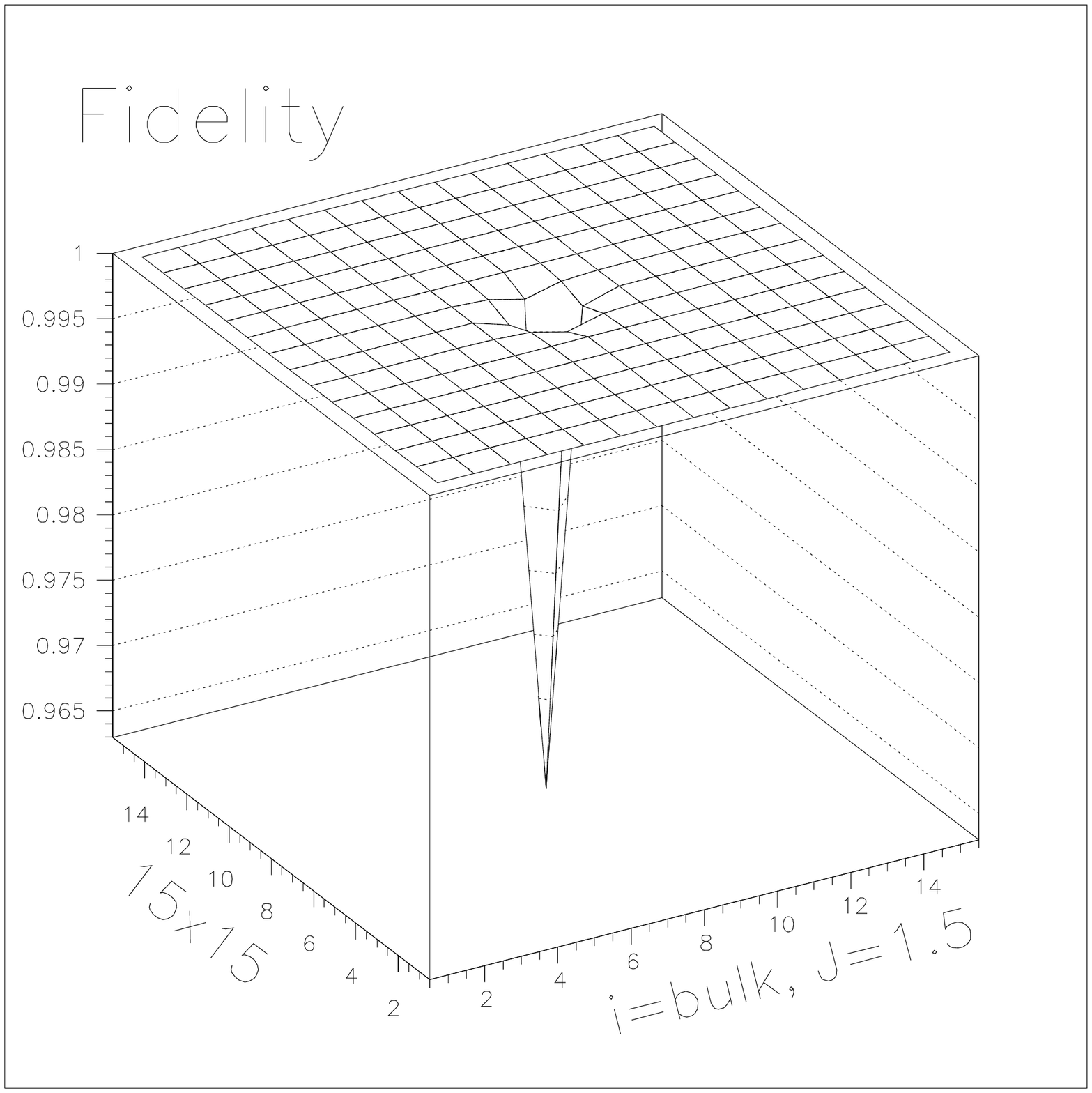}
\includegraphics[width=0.4\textwidth]{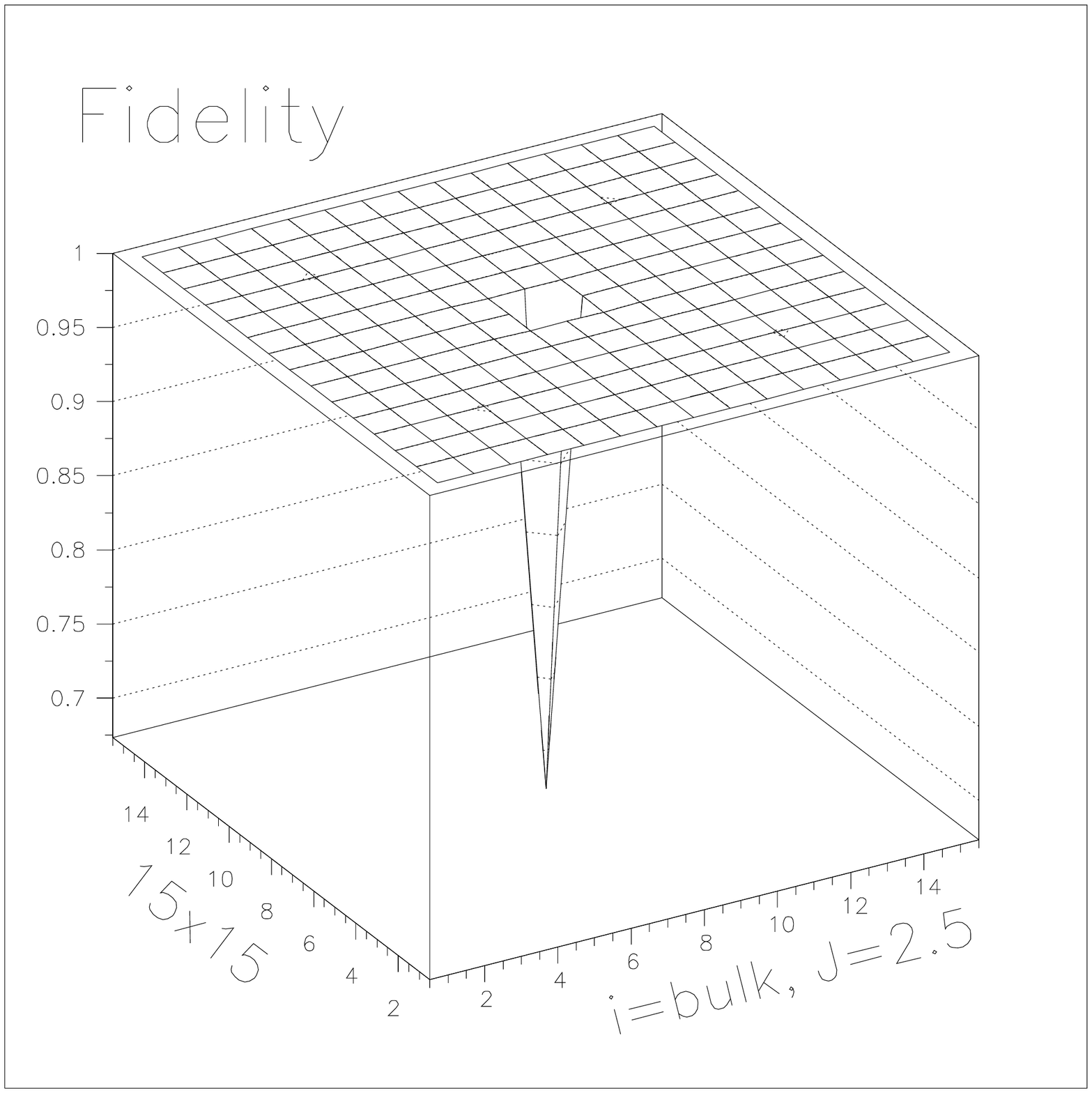}
\includegraphics[width=0.4\textwidth]{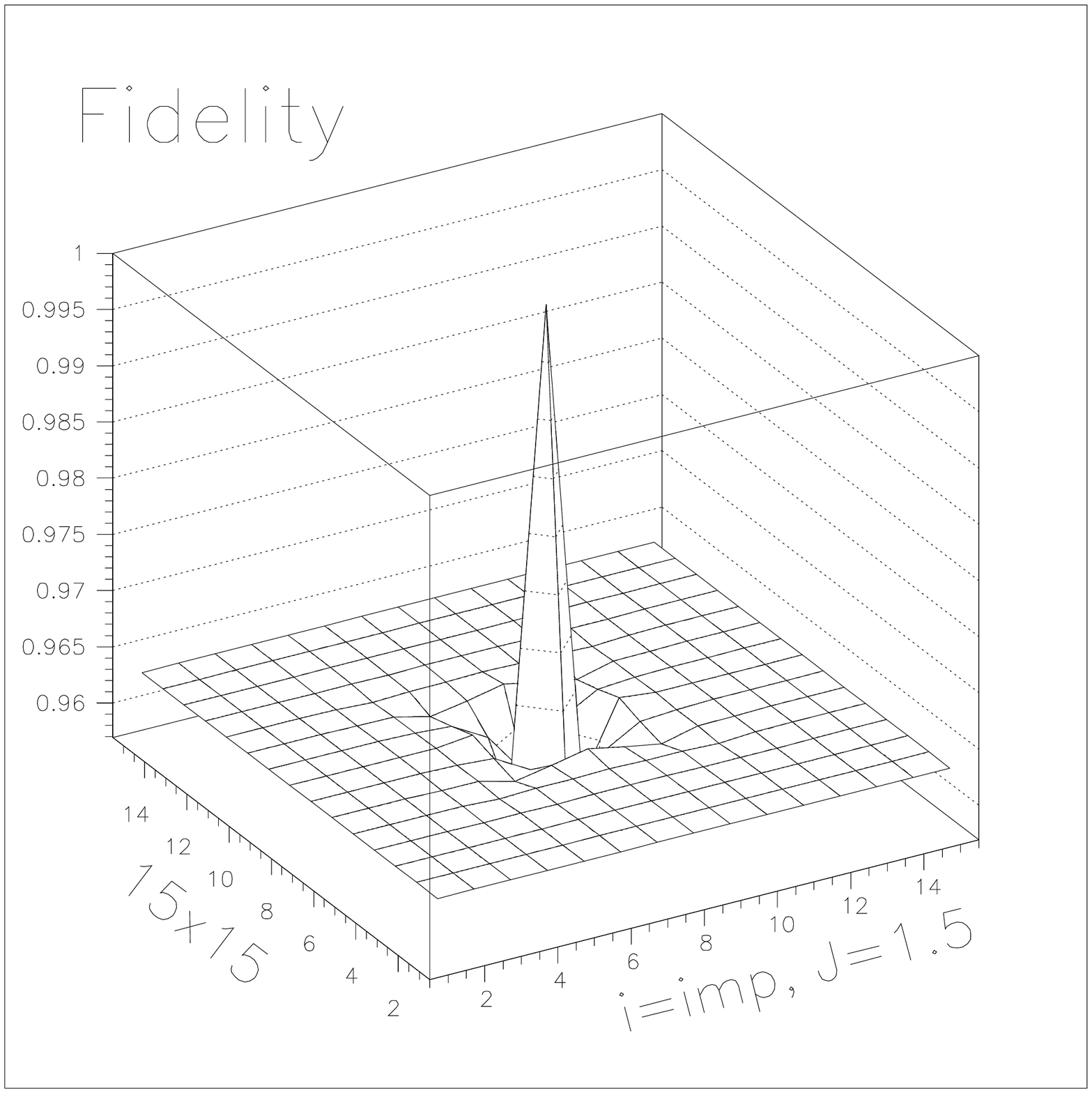}
\includegraphics[width=0.4\textwidth]{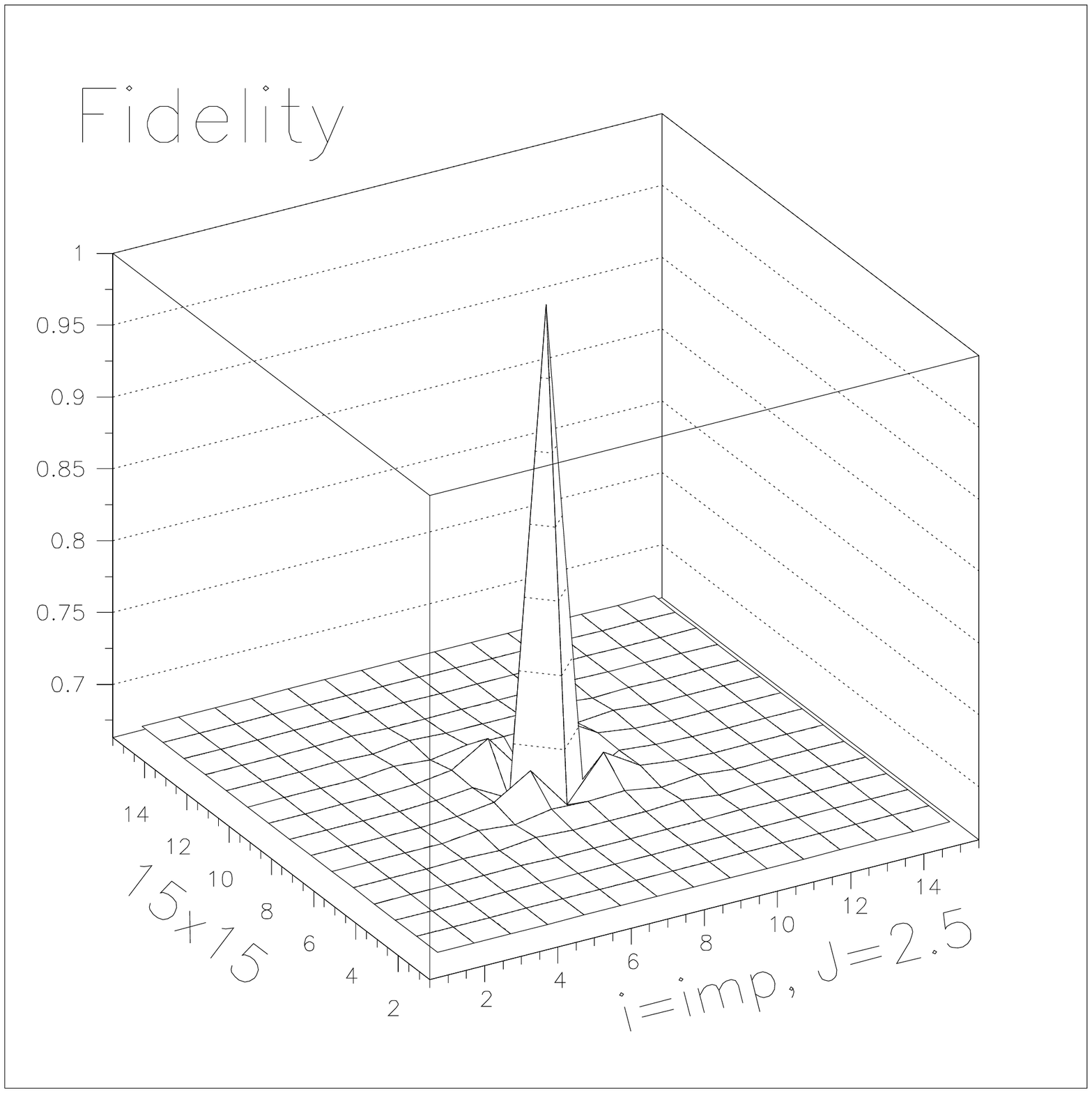}
\caption{\label{fig3} One-site fidelity, given for fixed values of
the parameter $J$, between any site and: a site in the bulk (upper
figures); a site in the impurity (lower figures). Left figures are
taken for $J^-=1.5<J_0$, figures on the right for $J^+=2.5>J_0$,
where $J_0=1.9$ is the point of the quantum phase transition. Note
the drop of the fidelity between the impurity and the bulk state,
from $F_{1}^-\simeq 0.96$ for $J^-=1.5<J_0$, to $F_{1}^+\simeq 0.65$
for $J^+=2.5>J_0$
--- another signature of the quantum phase transition.}
\end{figure*}

Finally, to further investigate the structure of one-site partial
states, we have evaluated the fidelity between the states for two
different sites, and the same value of the parameter $J$. The
results are presented in Fig. \ref{fig3}. Again, we see that apart
from the state of the site $l_c$ located at the impurity, and those
at its vicinity (first neighbors), all the other one-site partial
bulk states are almost the same, for a given common value of the
parameter $J$. In Fig. \ref{fig3} we have presented the results for
two distinct values of the parameter, $J^-=1.5<J_0$ and
$J^+=2.5>J_0$, but the qualitative picture stays the same for every
$J^-<J_0$ and $J^+>J_0$: the numerical value for the fidelity
between the bulk state (say, $\hat{\rho}_1$) and the impurity state
(say, $\hat{\rho}_2$) is significantly greater for $J^-$ than for
$J^+$, in other words,
$F^-_{1}(\hat{\rho}_a,\hat{\rho}_b)>F^+_{1}(\hat{\rho}_a,\hat{\rho}_b)$,
thus revealing the point of the quantum phase transition.

Note that when $i=l_c$ and $J<J_0$, besides the peak at the impurity
site, in its vicinity the fidelity $F_1$ is smaller than the
background value, while for $J>J_0$ it is larger. This is directly
connected to the existence of oscillations of the magnetization
around the impurity site (see \cite{first}).
Within the bounds set by quantum mechanics, two quantum states can
be distinguished by measuring different observables and comparing
the outcomes obtained. Magnetization is just one of them, and in
this case we see that it is optimal, or at least nearly so: for
$J<J_0$, it is easier to distinguish the impurity site from a site
in its immediate vicinity than from other, more distant sites,
simply because the difference in the expected magnetization between
two sites is bigger in the case of impurity-neighborhood than in
other cases. Therefore, in this case the fidelity has a drop in the
neighborhood of the impurity, ie the distinguishability is larger
there than elsewhere. A similar analysis holds for $J>J_0$.

\begin{figure}
%\begin{picture}(300,230)
%\put(-5,0){\includegraphics[width=0.70\textwidth]{Fig4}}
%\end{picture}
\includegraphics[width=0.70\textwidth]{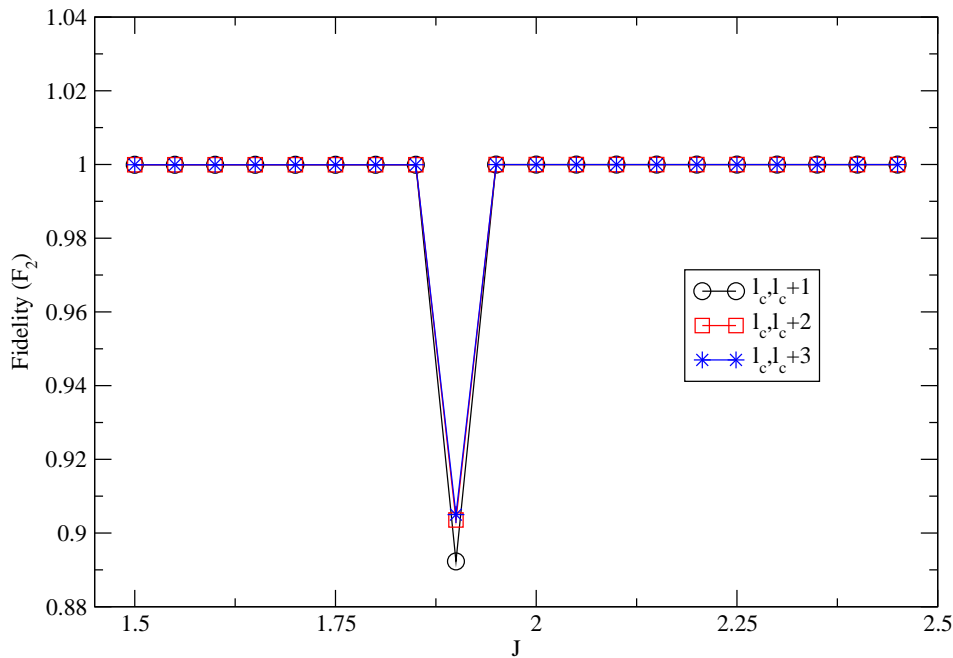}
\caption{\label{fig4} Two-site fidelity between two states taken for
two close parameter points $J$ and $J+\delta J$ ($\delta J=0.05$),
when one site is at the impurity and other in the neighborhood of
it. Note the sudden drop of the fidelity around the point of the
quantum phase transition, for the case of impurity $+$
neighboring site.}
\end{figure}

\subsection{Fidelity Between Two-site States}

We have also studied the behavior of the fidelity
taken between two-site partial states. We considered two distinct
cases. First, we calculated the fidelity between two-site states for
which one of the sites $i,j$ is at the impurity, say $i=l_c$, and
the other is in its vicinity, $j=l_c+\delta l$ ($\delta l=1,2,\ldots
6$), for two close parameter values $J$ and $J+\delta J$, with
$\delta J=0.05$. In Fig. \ref{fig4} we present the results for the
two-site fidelity $F_2(\hat{\rho}_a,\hat{\rho}_b)$, with
$\hat{\rho}_a=\hat{\rho}(J;l_c,l_c+\delta l)$ and
$\hat{\rho}_b=\hat{\rho}(J+\delta J;l_c,l_c+\delta l)$, where for
clarity we plotted only the $\delta l=1,2,3$ cases, as the other
curves almost overlap the results for $\delta l= 3$. Again, as in the previous
case of the fidelity between the one-site states, here as well the
sudden drop of the fidelity from its common value
$F_2(\hat{\rho}_a,\hat{\rho}_b)\simeq 1$ can be observed precisely
in the vicinity of the critical point $J_0=1.9$. We see it as a
signature of both the change in the impurity one-site partial
states, as well as the change of the total correlations between the
impurity and the remaining sites. From our study of the one-site
fidelity, we have seen that the bulk partial states are all
approximately the same, irrespectively of the site and the parameter
value, while only the impurity and first neighbor one-site states
exhibit a visible dramatic change around the point of the quantum
phase transition $J_0$.

\begin{figure}
\includegraphics[width=0.70\textwidth]{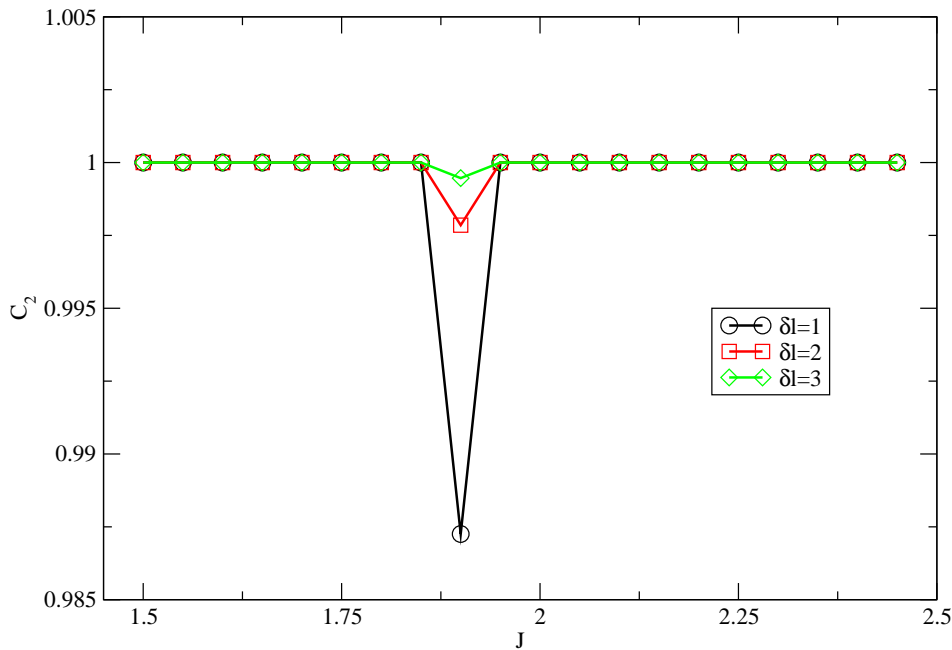}
\caption{\label{fig5} The $C_2$ function between two states taken
for two close parameter points $J$ and $J+\delta J$ ($\delta
J=0.05$), when one site is at the impurity and other in the
neighborhood of it. Note its sudden drop around the point of the
quantum phase transition, as well as the decrease of its deviation
from $1$ as the other site gets further from the impurity.}
\end{figure}

On the other hand, for every value of $J$ we can consider the quantity
\begin{equation}
\label{C_2} C_2(J,J+\delta J;l_c,l_c+\delta l)\equiv
\frac{F_2(\hat{\rho}(J;l_c,l_c+\delta l),\hat{\rho}(J+\delta
J;l_c,l_c+\delta l))}{F_1(\hat{\rho}(J;l_c),\hat{\rho}(J+\delta
J;l_c))F_1(\hat{\rho}(J;l_c+\delta l),\hat{\rho}(J+\delta
J;l_c+\delta l))},
\end{equation}
which can be seen as an indicator of total correlations, as in
the case of uncorrelated two-site composite systems the above
quotient reduces identically to $1$.
Therefore, the two-site composite system is necessarily correlated
whenever $F_2\neq 1$. Note that $C_2$
is {\em not} the measure of correlations --- the two-site composite
system can still exhibit nonzero correlations between its one-site
subsystems even for $F_2=1$. This is clearly seen in Fig.
\ref{fig5}, where $C_2=1$ for all values of $J$ except for $J=J_0$,
although is was shown in \cite{impurity-entanglement} that the
considered two-site systems are correlated for every value of $J$.
Yet, precisely around the point $J_0$ of the phase transition, the
quantity $C_2$ exhibits an abrupt change, thus showing an enhanced
change in the amount of the total correlations between the one-site
subsystems of an overall two-site system, see Fig. \ref{fig5}. We
also see that its numerical value around $J=J_0$ approaches $1$
as the step $\delta l$ increases --- ie the change of total
correlations decreases as the distance between the two sites
increases. This
behavior was already observed before in the study of total and
quantum correlations (entanglement) in this model
\cite{impurity-entanglement}. In connection to this, we note here
that the fidelity induces various distance measures on the set of
quantum states (for an overview, see for example \cite{fuchs} and
references therein), which can further induce fidelity-based
multi-particle entanglement measures, based on the quantum state
distinguishability, similarly to the case of the well known relative
entropy of entanglement measures \cite{relative_entropy}.

In the case of two-site subsystems with both one-site electrons
located in the bulk, the fidelity taken between states obtained for
two close parameter values $J$ and $J+\delta J$, is essentially
equal to one (with $\delta J=0.05$). Analogously to the case of the
fidelity between one-site partial states, here as well we see that
the impurity does not affect considerably the partial states of the
bulk of the lattice, even around the point of the quantum phase
transition. Yet, we have observed a small deviation around $J_0$ from the otherwise
common value $F_2\simeq 1$ which indicates that for larger subsystems
(when considering the fidelities $F_3$, $F_4$, etc. of the
$3$-electron and $4$-electron subsystems respectively), the impurity
might start to increasingly influence the collective behavior of the
bulk.

Finally, as in the previous case, we have evaluated the Uhlmann
mixed state geometric phase $\phi_g$, again obtaining the trivial
value $\phi_g\simeq 0$ everywhere, apart for the case when one
electron site is located at the impurity: the structure of the
two-site partial states' eigenvectors is not considerably affected
by the quantum phase transition.

\section{Conclusion}

In this paper we have applied the concept of fidelity to approach
quantum phase transitions, taking as the physical system a
conventional $2$-dimensional superconductor with a classical
magnetic impurity placed at its centre.
Unlike previous studies, where the overall states of the system
were considered, here we have studied the fidelity for partial
one- and two-site states.
We have shown that the point of the quantum phase transition is still
clearly marked by the sudden drop of the mixed state fidelity, even
in the simplest case of the one-site partial states, as long as they
are close to the impurity itself.
The same holds for the two-site partial states --- the fidelity
exhibits a sudden drop in the vicinity of the quantum phase transition.
Here, for the transition to be inferred it is necessary that one of the
subsystem electrons lies at the impurity, or its neighborhood.
This is expected to occur for this model since, as shown earlier,
the impurity acts mainly as a local perturbation (but see also
the behavior of the non vanishing two-site ``long distance''
entanglement measures discussed in \cite{impurity-entanglement}).
We have also evaluated the Uhlmann mixed state geometric phase for
the cases of both one- and two-site partial states and found it to
be trivial (ie equal to one) for every value of the quantum phase
transition driving parameter $J$, except when an electron site is
at the impurity, thus showing that the structure of the subsystem
eigenvectors has little relevance to the magnetic transition discussed.
Finally, we have shown that the fidelity
approach can also be useful in studying the total correlations
between subsystems of the whole physical system, and how the
point of quantum phase transition can be inferred from the change
in the amount of total correlations (given by the quantity $C_2$
(\ref{C_2})), which is closely related to the fidelity between the
partial states discussed in this work.

We believe that this work offers yet another confirmation of the
usefulness of the fidelity approach to study quantum phase transitions.
It extends this approach to partial states, showing that, in some cases
at least, quantum phase transitions can be signaled out by the
behavior of the mixed state fidelity taken between local partial
states. It is an interesting and challenging task to try to find a
unified view to this new approach and possibly connect it to some
previously well studied thermodynamic quantities. Also, the
generalization to thermal phase transitions is another possible path
of further research. Finally, we see our simplified study of the
total correlations in a composite quantum state, based on the
fidelity, as just a first step towards exploring the possibility of
using the fidelity-based quantum state distinguishability for
quantifying correlations, both classical as well as quantum
(entanglement), a subject of research that has on its own found rich
applications within the field of condensed matter and many-body
physics.

\section{Acknowledgments}

NP thanks the support from Funda\c{c}\~{a}o para a Ci\^{e}ncia e a
Tecnologia (Portugal), grant SFRH/BPD/31807/2006. PDS, PN and VRV
acknowledge the support of projects SFA-2-91, the ESF Science Programme INSTANS 2005-2010, POCI/FIS/58746/2004
and FCT POCI/MAT/55796/2004. VKD is supported by the Polish
Ministry of Science and Higher Education as research projects in years
2006-2009 and 2007-2010 and by STCU Grant No. 3098 in Ukraine. This work
is also supported by the cooperation agreement between Poland and
Portugal in the years 2007-2008.

%______________________________________________________________________ BIBLIOGRAPHY

\end{document}